\documentclass[12pt,a4paper]{article}
\usepackage{a4wide}
\usepackage{amssymb}
\usepackage{graphicx}
\usepackage{cite}
\begin{document}
\Large
\begin{center}
\textbf{Multiplicative point process as a model of trading activity}
\\[1.25\baselineskip]
\small \textbf {V. Gontis and B. Kaulakys\footnote{Corresponding author.
Tel. +370-5-2620859; fax +370-5-2125361.
\\
\textit{E-mail address}: kaulakys@itpa.lt (B. Kaulakys)}}

\vskip 2mm Institute of Theoretical Physics and Astronomy,
Vilnius University \\ A.Go\v{s}tauto 12, LT-01108 Vilnius, Lithuania \\
gontis@ktl.mii.lt
\end{center}

\abstract{Signals consisting of a sequence of pulses show that
inherent origin of the $1/f$ noise is a Brownian fluctuation of
the average interevent time between subsequent pulses of the pulse
sequence. In this paper we generalize the model of interevent time
to reproduce a variety of self-affine time series exhibiting power
spectral density $S(f)$ scaling as a power of the frequency $f$.
Furthermore, we analyze the relation between the power-law
correlations and the origin of the power-law probability
distribution of the signal intensity. We introduce a stochastic
multiplicative model for the time intervals between point events
and analyze the statistical properties of the signal analytically
and numerically. Such model system exhibits power-law spectral
density $S(f)\sim1/f^\beta$ for various values of $\beta$,
including $\beta=1/2$, 1 and 3/2. Explicit expressions for the
power spectra in the low frequency limit and for the distribution
density of the interevent time are obtained. The counting
statistics of the events is analyzed analytically and numerically,
as well. The specific interest of our analysis is related with the
financial markets, where long-range correlations of price
fluctuations largely depend on the number of transactions. We
analyze the spectral density and counting statistics of the number
of transactions. The model reproduces spectral properties of the
real markets and explains the mechanism of power-law distribution
of trading activity. The study provides evidence that the
statistical properties of the financial markets are enclosed in
the statistics of the time interval between trades. A
multiplicative point process serves as a consistent model
generating this statistics.}

\section{Introduction}

\label{intro} Complex collective phenomena usually are responsible
for the power-laws which are universal and independent of the
microscopic details of the phenomenon. Examples in physics are
numerous. Power-laws are intrinsic features of the economic and
financial data, as well. The aim of this contribution is to
analyze a relationship between the origin of the power-law
distribution and the power-law correlations in financial time
series.

There are numerous studies of power-law probability distributions
in various economic systems \cite{1,1a,2,3,3a,3b,4,4a,41,42,8,8a,5,5a,5b}.
The key result in recent findings is that the cumulative distributions of
returns and trading activity can be well described by a power-law
asymptotic behavior, characterized by an exponent $\lambda \approx
3$, well outside the stable Levy regime $0<\lambda <2$ \cite {5,5a,5b}.
Empirical studies confirm the power-law asymptotic behavior with stable
exponent for different time scales. Since these empirical distributions
are neither Levy stable nor invariant under addition, new stochastic models
with long range correlations and power-law asymptotic behavior are of great
interest \cite{5a}.

The random multiplicative process built into the model of wealth
distribution yields Pareto power-law \cite{1a}. The
generalized Lotka-Volterra dynamics is in the use for various
systems including financial markets \cite{41}. However, these
models generically lead to non-universal exponents and do not
explain the power-law autocorrelations and observable spectral density of
financial time series \cite{8}. The time-correlations in the financial
time series are studied extensively as well \cite{5,5a,5b,6,7}.
Recent investigations \cite{5,5a,5b,6} provide empirical evidence
that the long-range correlations for volatility are due to the trading
activity, measured by a number of transactions $N$. Therefore, various
stochastic models of trading activity or for waiting time between
transactions have to be analysed for reproduction the statistics of
volatility in financial markets.

Recently we tried to adapt the model of $1/f$ noise based on the Brownian
motion of time interval between subsequent pulses, proposed in
\cite{9,10,11,12}, to model the share volume
traded in the financial markets \cite{13}. The idea to transfer
long time correlations into the stochastic process of the time
interval between trades or time series of trading activity is in
consistence with the detailed studies of the empirical financial
data \cite{5,6} and can reproduce the spectral properties of
the financial time series \cite{13}. However, the investigation of
the model revealed, that the simple additive Brownian model of
time interval between trades failed to reproduce the power-law
probability density function (pdf) of the trading activity
\cite{131}.

On the other hand, several authors have shown
empirically that the fluctuations of various financial time series
possess multifractal statistics \cite{14,15,16,17,18}.
Therefore, in this paper we introduce the stochastic multiplicative
model for the time interval between trades and analyze the statistical
properties of the trading activity analytically and numerically.
The model is in consistence with the results of statistical analysis of the
empirical financial time series and reveals the relationship
between the power-law probability distribution and the power-law
spectral density of the financial time series. Moreover,
the analytically solvable model \cite{9,10,11,12} results only in exact
$1/f$ noise while in the papers \cite{13,131} essentially only numerical analysis
has been undertaken. Here we generalize the model defining the time series
exhibiting the power spectral density $1/f^{\beta }$ with $
0.5\lesssim \beta \lesssim 1.5$ and the power-law probability density
of the trading activity. Explicit analytical expressions for the power
spectra and for pdf of the trading activity $N$ are obtained and analysed.

\section{The model}

We consider a signal $I(t)$ as a sequence of the random correlated pulses
\begin{equation}
\label{eq:1}I(t)=\sum\limits_ka_k\delta (t-t_k)
\end{equation}
where $a_k$ is a contribution to the signal of one pulse at the
time moment $t_k$, e.g., a contribution of one transaction
to the financial data. Signal (\ref{eq:1}) represents a point
process used in a large variety of systems with the flow of
point objects or subsequent actions. When $a_k=\bar a$ is
constant, the point process is completely described by the set of
times of the events $\{t_k\}$ or equivalently by the set of
interevent intervals $\{\tau _k=t_{k+1}-t_k\}$. Various stochastic
models of $\tau _k$ can be introduced to define a stochastic point
process. In papers \cite{9,10,11,12} it has been shown
analytically that the relatively slow Brownian fluctuations of the
interevent time $\tau _k$ yield $1/f$ fluctuations of the signal
(\ref{eq:1}).

Power spectral density of the signal $I(t)$ is
\begin{eqnarray}
S(f)&=&\lim_{T\rightarrow \infty }\left\langle \frac{2\overline{a}^{2}}{T}
\left| \int\limits_{t_{i}}^{t_{f}}I(t)\exp (-i2\pi ft)dt\right|^{2}
\right\rangle \label{eq:2}  \\
&=&\lim_{T\rightarrow \infty }\left\langle \frac{2\overline{a}^{2}}{T}
\sum_{k=k_{\min }}^{k_{\max }}\sum_{q=k_{\min }-k}^{k_{\max
}-k}\exp (-i2\pi f\Delta (k;q))\right\rangle \nonumber
\end{eqnarray}
where the brackets $\left\langle ...\right\rangle $ denote the averaging
over the realizations of the process, $T=t_f-t_i$ is the observation time,
$\Delta (k;q)=t_{k+q}-t_k$ is the time difference between the pulses
occurrence times $t_{k+q}$ and $t_k$, while $k_{\min }$ and $k_{\max }$ are
minimal and maximal values of the index $k$ in the interval of observation
$T $.

We adapt and generalize the model of $1/f$ noise previously
proposed in \cite {9,10,11,12} to model the share volume traded in
the financial markets \cite {5,6}. It is useful to define a
discrete time series with equal time intervals $\tau _d$.
Integration of the signal $I(t)$ over subsequent time intervals of
length $\tau _d$ results in a discrete time series and, by analogy
with financial time series, we will call it the volume $V_j$,
\begin{equation}
\label{eq:3}V_j=\int\limits_{t_j}^{t_j+\tau
_d}I(t)dt=\sum\limits_{t_j<t_k<t_j+\tau _d}a_k,\quad t_j=j\tau _d.
\end{equation}

The number of trades $N_j$ in the time interval $[t_j, t_j+\tau _d]$ is
defined by the same equation (\ref{eq:3}) with $a_k\equiv 1$ and $N_j\equiv
V_j$. Then the power spectral density $S(f_s)$ of the discrete signal $N_j$
may be calculated by the FFT as
\begin{equation}
\label{eq:4}S(f_s)=\left\langle \frac 2{\tau _dn}\left| \sum_{j=1}^nN_j\exp
(-i2\pi (s-1)(j-1))\right| ^2\right\rangle.
\end{equation}
Here the discrete frequencies $f_s=(s-1)/T$, $s=1, 2,...n$ and $T=\tau _dn$.
For frequencies $f_s\ll 1/\tau _d$ the power spectral density defined by
equation~(\ref{eq:4}) coincides with that given by equation~(\ref{eq:2}).

In this paper we investigate the statistical properties of the
time series $N_j$, when the sequence of interevent time $\tau _k$
is generated by a multiplicative stochastic process. First of all
the multiplicativity is an essential feature of processes in
economics \cite{1,4}. Multiplicative stochastic processes yield
multifractal intermittency and are able to produce power-law
probability distribution functions. We base our study on the
generic multiplicative process for the interevent times, written
as
\begin{equation}
\label{eq:5}\tau _{k+1}=\tau _k+\gamma \tau _k^{2\mu -1}+\tau _k^\mu \sigma
\varepsilon _k.
\end{equation}
Here the interevent time $\tau _k$ fluctuates due to the external random
perturbation by a sequence of uncorrelated normally distributed random
variable $\{\varepsilon _k\}$ with zero expectation and unit variance,
$\sigma $ denotes the standard deviation of the white noise and $\gamma \ll 1$
is a damping constant.

We will restrict the diffusion of the interevent time according to
equation (\ref{eq:5}) to the finite interval $\left[ \tau _{\min
},\tau _{\max }\right] $, i.e., $0<\tau _{\min }<\tau _k<\tau
_{\max }$. The most simple case is the pure multiplicativity of
$\tau _k$, i.e., when $\mu =1$. Other values of $\mu $ reproduce
the power-laws as well and explicit expressions can be derived
without the loss of generality.

The iterative relation (\ref {eq:5}) can be rewritten as Langevine
stochastic differential equation in $k$-space.
\begin{equation}
\label{eq:6}\frac{d\tau _k}{dk}=\gamma \tau _k^{2\mu -1}+\tau _k^\mu \sigma
\xi \left( k\right).
\end{equation}
Here we interpret $k$ as continuous variable while
$$
\left\langle \xi \left( k\right) \xi \left( k^{\prime }\right)
\right\rangle =\delta (k-k^{\prime }).
$$

The steady state solution of the corresponding stationary
Fokker-Planck equation with zero flow gives the probability
density function for $\tau _k$ in the $k$-space (see, e.g.,
\cite{19})
\begin{equation}
\label{eq:7}P_k(\tau _k)=C\tau _k^\alpha ,\quad \alpha =2\gamma /\sigma
^2-2\mu
\end{equation}
and $C$ has to be defined from the normalization
$$
\int_{\tau _{\min }}^{\tau _{\max }}P_k(\tau )d\tau =1.
$$

The solution (\ref{eq:7}) assumes Ito convention involved in the relation
between expressions (\ref{eq:5}), (\ref{eq:6}) and (\ref{eq:7}).

As the probability distribution function (\ref{eq:7}) follows a
power-law, we expect that this would result in the power-law
behavior in other statistics, as well. In the next Section we will
obtain the power spectral density defined by equation
(\ref{eq:2}).

\section{Power spectral density}

The power spectral density is a well-established measure of long
time correlations and is widely used in stochastic systems. As it
has been already shown \cite{9,10,11,12} the point process with
the Brownian interevent time exhibits $1/f$ noise.

We will derive the formula for the power spectral density of the
multiplicative stochastic point process model, defined by equations~(\ref
{eq:5}) and (\ref{eq:6}) for the interevent time.

Let us rewrite equation~(\ref{eq:2}) as \cite{12}
\begin{equation}
\label{eq:8}S(f)=\lim _{T\rightarrow \infty }\frac{2\overline{a}^2}
T\sum\limits_{k,q}\chi _{\Delta (k:q)}(2\pi f)
\end{equation}
where $\chi _{\Delta (k:q)}(2\pi f)$ is the characteristic function for the
probability distribution of $\Delta (k;q)$. For the normal distribution of
$\Delta (k;q)$ formula (\ref{eq:8}) takes the form
\begin{equation}
\label{eq:9}S(f)=\lim _{T\rightarrow \infty }\frac{2\overline{a}^2}
T\sum\limits_{k,q}\exp \{i2\pi f\left\langle \Delta
(k;q)\right\rangle -2\pi ^2f^2\sigma _\Delta ^2(k;q)\}
\end{equation}
where $\sigma _\Delta ^2$ is the variance of $\Delta (k;q)$, the time
difference between the pulses occurrence times $t_{k+q}$ and $t_k$. For $\mu
=1$ the time difference $\Delta (k;q)$ may be expressed from the solution of
multiplicative stochastic equation~(\ref{eq:6}) as
\begin{equation}
\label{eq:10}\Delta (k;q)=\sum\limits_{l=1}^q\tau _k\exp \{
(\gamma -\frac 12\sigma ^2)l+\sigma \sum\limits_{j=1}^l\varepsilon _j\}.
\end{equation}

Averaging over the normal distributions of uncorrelated $\varepsilon _j$
results in explicit expressions for the mean $\left\langle \Delta
(k;q)\right\rangle $ and variance $\sigma _\Delta ^2(k;q)$.

In general, for any $\mu $, the perturbative solution of equation (\ref{eq:6}
) yields
\begin{eqnarray}
\left\langle \Delta (k;q)\right\rangle &=&\tau _{k}q+\frac{\gamma
}{2}\tau
_{k}^{2\mu -1}q^{2}+o(\gamma ^{2}), \label{eq:11} \\
\sigma _{\Delta }^{2}(k;q) &=&\tau _{k}^{2\mu }\frac{\sigma
^{2}}{3}q^{3}. \nonumber
\end{eqnarray}

In the low frequency limit $f\ll \tau _k^{-1}$ we can replace the
summation over $k$ and $q$ by the integration and take into
account only the first order terms of equation (\ref{eq:11}) in
the expression for the power spectral density (\ref {eq:9}). This
yields
\begin{eqnarray}
S_{\mu }(f)&=&\frac{4C\overline{a}^{2}}{\overline{\tau }}
{\mathop{\mathrm{Re}}}\int\limits_{\tau _{\min }}^{\tau _{\max }}
d\tau\tau ^{\alpha }\int\limits_{0}^{\infty }
\exp \{i2\pi f(\tau q+\frac{\gamma}{2}\tau ^{2\mu
-1}q^{2})\}dq, \nonumber \\
&=&\frac{2C\overline{a}^{2}}{\sqrt{\pi }\overline{\tau }(3-2\mu
)f}\left(\frac{\gamma }{\pi f}\right)^{\frac{\alpha }{3-2\mu
}}I_{\mathrm{erf}}(x_{\min },x_{\max }), \label{eq:12}
\end{eqnarray}
\begin{eqnarray}
I_{{\mathop{\mathrm{erf}}}}(x_{\min },x_{\max })&=&
{\mathop{\mathrm{Re}}}\int\limits_{x_{\min }}^{x_{\max }}\exp
\left\{-i\bigg(x-\frac \pi 4\bigg)\right\}{\mathop{\mathrm{erfc}}}
(\sqrt{-ix})x^{\frac\alpha {3-2\mu }-\frac 12}dx \label{eq:13}
\end{eqnarray}
where
$$
\bar \tau =\left\langle \tau _k\right\rangle =\frac T{k_{\max
}-k_{\min }}
$$
is the expectation of $\tau _k$. Here we introduce the scaled variable $x=\frac{
\pi f}\gamma \tau ^{3-2\mu }$ and
$$
x_{\min }=\frac{\pi f}\gamma \tau _{\min }^{3-2\mu },\quad x_{\max }=\frac{
\pi f}\gamma \tau _{\max }^{3-2\mu }.
$$

Expression of the power spectral density (\ref{eq:12}) is
appropriate for the numerical calculations of the generalized
multiplicative point process defined by equations (\ref{eq:1}) and
(\ref{eq:5}). In the limit $\tau _{\min }\rightarrow 0$ and $\tau
_{\max }\rightarrow \infty $ we obtain an explicit expression for
$S_\mu (f)$
\begin{equation}
\label{eq:14}S_\mu (f)=\frac{C\overline{a}^2}{\sqrt{\pi }\overline{\tau }
(3-2\mu )f}\left(\frac \gamma {\pi f}\right)^{\frac \alpha {3-2\mu
}}\frac{\Gamma (\frac 12+\frac \alpha {3-2\mu })}{\cos (\frac{\pi
\alpha }{2(3-2\mu )})}.
\end{equation}

Equation (\ref{eq:14}) reveals that the multiplicative point
process (\ref{eq:5}) results in the power spectral density
$S(f)\sim 1/{f^\beta }$ with the scaling exponent
\begin{equation}
\label{eq:15}\beta =1+\frac{2\gamma /\sigma ^2-2\mu }{3-2\mu }.
\end{equation}

\begin{figure}[htb]
\begin{center}
\includegraphics{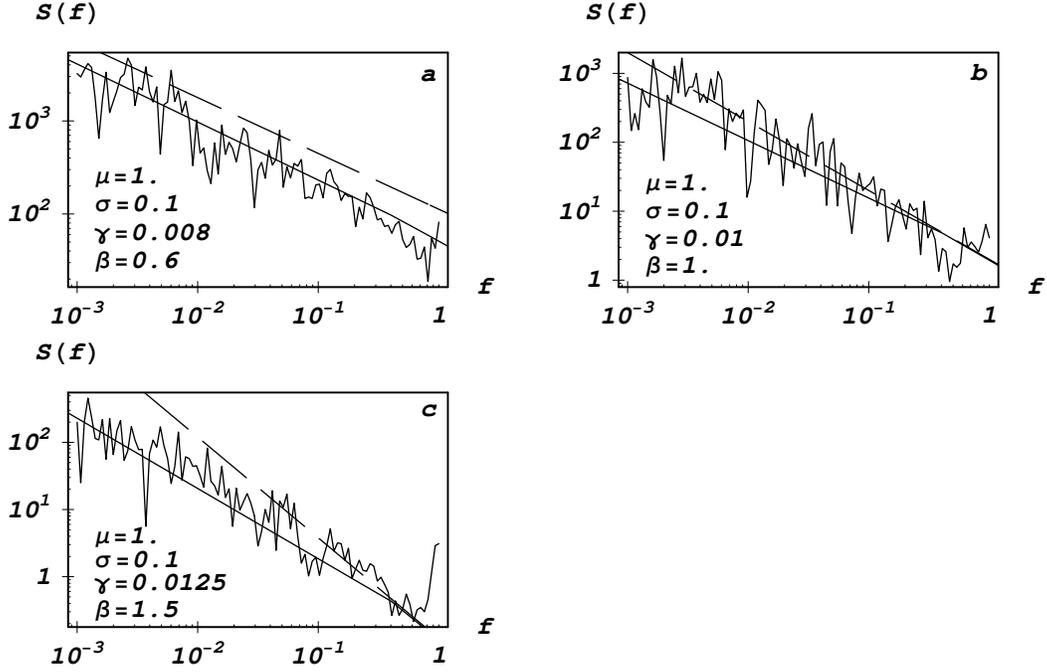}
\end{center}
\caption{The power spectral density $S(f)$ of the signal $I(t)$
versus frequency $f$ calculated from the model described by
equations (\ref{eq:1}), (\ref{eq:2}) and (\ref{eq:5}) with
$\mu=1$. The sinous curves represent the results of the numerical
simulations averaged over 5 realizations of 100000 impulse
sequences with $\sigma=0.1$, the straight lines represent
the analytical approximation given by integral formula (\ref{eq:12})
with $\tau _{min}=10^{-6}$; $\tau _{max}=1$ and the dashed lines
show results of explicit formula (\ref{eq:14}). (a) $\gamma=0.008$, (b)
$\gamma=0.01$ and (c) $\gamma=0.0125$.} \label{fig:1}
\end{figure}

\begin{figure}[htb]
\begin{center}
\includegraphics{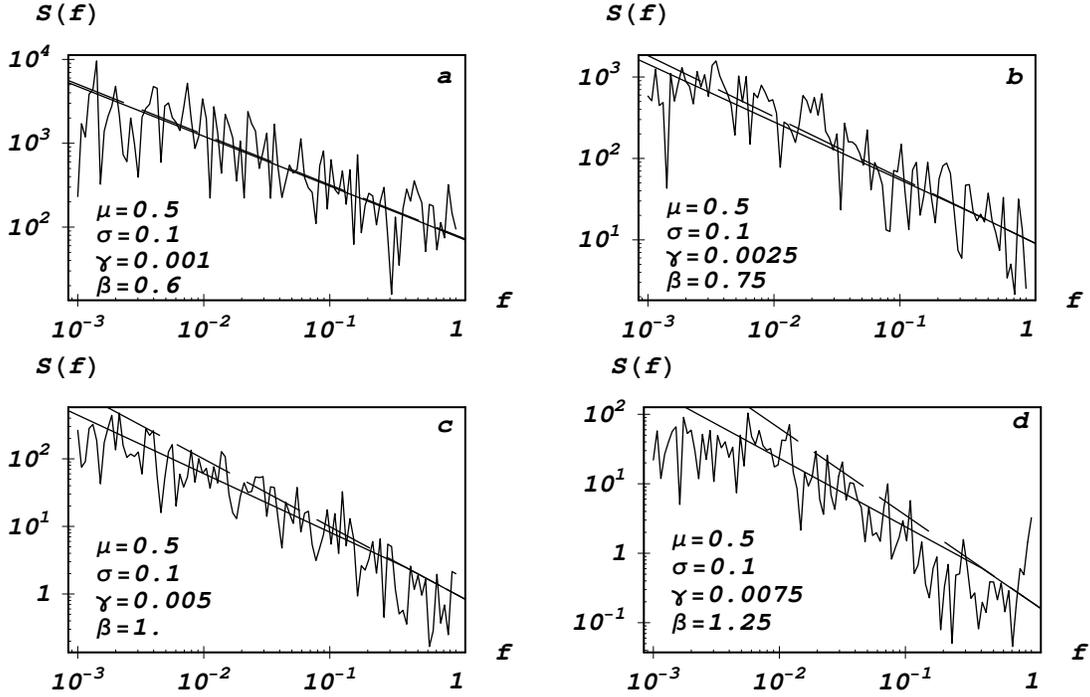}
\end{center}
\caption{The same as in Fig.~\ref{fig:1} but for $\mu=0.5$. (a)
$\gamma=0.001$, (b) $\gamma=0.0025$, (c) $\gamma=0.005$ and (d)
$\gamma=0.0075$.}

\label{fig:2}
\end{figure}

Let us compare our analytical results (\ref{eq:12}), (\ref{eq:13})
and (\ref{eq:14}) with the numerical calculations of the power
spectral density according to equations (\ref{eq:2}) and
(\ref{eq:5}). In Fig.~\ref{fig:1} we present the numerically
calculated power spectral density $S(f)$ of the signal $I(t)$ for
$\mu =1$ and $\alpha =2\gamma/\sigma ^2-2=0$, -0.4 and
+0.5. Numerical results confirm that the multiplicative point process
exhibits the power spectral density scaled as $S(f)\sim 1/f^\beta
$. Equation (\ref{eq:12}) describes the model power spectral
density very well in a wide range of parameters. The explicit
formula (\ref{eq:14}) gives a good approximation of power spectral
density for the parameters when $\beta \simeq 1$. In
Fig.~\ref{fig:2} we present numerical results of the power
spectral density calculated for the multiplicative point process with
$\mu =0.5$. In this case analytical expressions (\ref{eq:12}),
(\ref{eq:13}) and (\ref {eq:14}) describe power spectral density
very well in a wide range of the parameter $\gamma $. These
results confirm the earlier finding \cite {9,10,11,12} that the power
spectral density is related to the probability distribution of the
interevent time $\tau _k$ and $1/f$ noise occurs when this
distribution is flat, i.e., when $\alpha =0$.

It is likely that such a stochastic model with parameters in the
region $0.5\leq \beta \leq 1.5$ may be adaptable for a wide
variety of different systems. In this paper we will investigate
applicability of the model for the financial market.

\section{Power-law probability distribution function}

The statistics of the trading activity $N$, i.e., the number of trades
per time interval $\tau _d$ is the major task of this paper. The
consistency of any model of financial markets depends on its
ability to reproduce empirically measured probability distribution
of the number of trades $N$, defined by equation (\ref{eq:3}).
Intuitively, the pdf of $N$ is related with pdf of the rate
$\nu = 1/\tau$ and with pdf of $\tau $.
The pdf of $\tau _k$ in $k$-space is given by
expression (\ref{eq:7}). It is obvious that in
the actual time $t$ the pdf of $\tau $ may be written as
\begin{equation}
\label{eq:16}P_t(\tau )=P_k(\tau )\tau /\bar \tau =C^{^{\prime }}\tau
^{\alpha +1}
\end{equation}
where $C^{^{\prime }}=C/\bar \tau $ is a new normalization
constant.

For the pure multiplicative model with $\mu =1$
equations~(\ref{eq:10}) and (\ref {eq:11}) define a relationship
between $N$ and $\tau $ after the
substitution $\left\langle \Delta (k;q)\right\rangle \rightarrow \tau _d$,
$\tau _k\rightarrow \tau $ and $q\rightarrow N$, i.e.,
\begin{equation}
\label{eq:17}\tau _d=\tau N+\frac \gamma 2\tau N^2.
\end{equation}

This relationship may be used for definition of pdf $P(N)$
from the relation \\$P(N)dN = P_t(\tau )d\tau $. We have
\begin{equation}
P(N) =\frac{C'\tau_{d}^{2+\alpha}(1+\gamma
N)}{N^{3+\alpha}(1+\frac{\gamma}{2}N)^{3+\alpha}}.
\label{eq:18} \\
\end{equation}

This equation yields limiting cumulative distributions for $N$,
\begin{equation}
P_{>}(N)=\int\limits_N^{+\infty}P(N)dN\sim\left\{
\begin{array}{ll}
\frac{1}{N^{2+\alpha}},& N\ll \gamma^{-1}, \\
\frac{1}{N^{4+2\alpha}},& N\gg \gamma^{-1}.
\end{array}
\right.
\label{eq:19}
\end{equation}

\begin{figure}
\begin{center}
\includegraphics{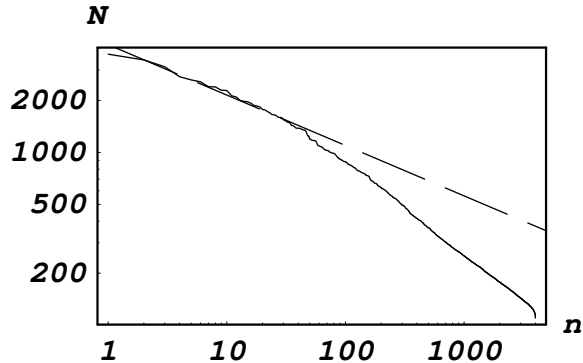}
\end{center}
\caption{Amplitude versus rank, $N(n)$, plot defining cumulative
distribution $P_{>}(N)\sim1/N^{\lambda}$ with the
parameters of the model $\mu=1$, $\sigma=0.1$ and $\gamma=0.0085$.
Dashed line represents asymptotic plot $N(n)\sim n^{- 1/
\lambda}$ with $\lambda=3.4$.}
\label{fig:3}       
\end{figure}

Let us compare probability distribution function for $N$ obtained from
the numerical simulation of the model with the analytical results
(\ref{eq:18}) and (\ref{eq:19}). It is convenient to use the amplitude
versus rank plot (see, e.g., \cite{20} ) for definition of the exponent
$\lambda$ of the cumulative distribution $P_{>}(N)\sim1/N^\lambda $. We
plot the value of the $n$-th rank variable $N(n)$ as a function
of its rank $n$.
If $P_{>}(N)$ behaves asymptotically when $n\rightarrow 0$ as
a power-law, one gets a straight line in log-log coordinates of
$N$ and $n$, with the slope equal to $-1/\lambda $. In
Fig.~\ref{fig:3} we present the plot for $N(n)$ calculated
numerically for pure multiplicative model ($\mu =1$) with the
parameters yielding the best fit to the empirical data.

\section{Discussion and conclusions}

We have introduced a multiplicative stochastic model for the time
intervals between events of point process. Such a model of time
series has only a few parameters defining the statistical
properties of the system, i.e., the power-law behavior of the
distribution function and the scaled power spectral density of the
signal. The ability of the model to simulate $1/f$ noise as well
as to reproduce signals with the values of power spectral density
slope $\beta$ between 0.5 and 1.5 promises a wide variety of
applications of the model.

Let us present shortly the possible interpretations of the
empirical data of the trading activity in the financial markets. With
a very natural assumption of transactions in the financial markets
as point events we can model the number of transactions $N_j$ in
equal time intervals $\tau _d$ as the outcome of the described
multiplicative point process. We already know from available
studies \cite{5} that the empirical data exhibit power spectral
density in the low frequency limit with the slope $\beta \simeq
0.7$. For the pure multiplicative model $\mu =1$ this corresponds
to the case when $\alpha =2\gamma /\sigma ^2-2\mu \simeq -0.3$.
The corresponding cumulative distribution of $N$ in the tail of
high values (see equation (\ref{eq:19})) has the exponent
$\lambda=4+2\alpha =3.4$. This is in an excellent agreement with
the empirical cumulative distribution exponent 3.4 defined in
\cite{5} for 1000 stocks of the three major US stock markets.

The numerical results confirm that the multiplicative stochastic
model of the time interval between trades in the financial market
is able to reproduce the main statistical properties of trading
activity $N$ and its power spectral density. The power-law
exponents of the pdf of the interevent time, $\alpha $, and the
cumulative distribution of the trading activity, $\lambda $, as
well as the slope of power spectral density, $\beta$, are defined
just by one parameter of the model $2\gamma /\sigma ^2$. The model
suggests a simple mechanism of the power-law statistics of trading
activity in the financial markets. We expect that
multiplicative model of the time interval between the trades with
more specific restrictions for the diffusion and more precisely adjusted
parameters lies in the background of the financial markets
statistics and may be useful in the financial time series
analysis.

\noindent \textbf{Acknowledgements}

We acknowledge support by the Lithuanian State and Studies Foundation.

\end{document}